\documentstyle[12pt]{article}
\begin{document}
\def\QATOPD#1#2#3#4{{#3 \atopwithdelims#1#2 #4}}
\def\stackunder#1#2{\mathrel{\mathop{#2}\limits_{#1}}}
\def\bea{\begin{eqnarray}}
\def\eea{\end{eqnarray}}
\def\nn{\nonumber}
\def\baselinestretch{1.5}
\def\beq{\begin{equation}}
\def\eeq{\end{equation}}
\def\ba{\beq\new\begin{array}{c}}
\def\ea{\end{array}\eeq}
\def\be{\ba}
\def\ee{\ea}
\def\stackreb#1#2{\mathrel{\mathop{#2}\limits_{#1}}}
\def\Tr{{\rm Tr}}
\def\res{{\rm res}}
\def\f{1\over}
\makeatletter
\newdimen\normalarrayskip              
\newdimen\minarrayskip                 
\normalarrayskip\baselineskip
\minarrayskip\jot
\newif\ifold             \oldtrue            \def\new{\oldfalse}
\def\arraymode{\ifold\relax\else\displaystyle\fi} 
entries
\def\eqnumphantom{\phantom{(\theequation)}}     
eqnarray
\def\@arrayskip{\ifold\baselineskip\z@\lineskip\z@
     \else
     \baselineskip\minarrayskip\lineskip2\minarrayskip\fi}
\def\@arrayclassz{\ifcase \@lastchclass \@acolampacol \or
\@ampacol \or \or \or \@addamp \or
   \@acolampacol \or \@firstampfalse \@acol \fi
\edef\@preamble{\@preamble
  \ifcase \@chnum
     \hfil$\relax\arraymode\@sharp$\hfil
     \or $\relax\arraymode\@sharp$\hfil
     \or \hfil$\relax\arraymode\@sharp$\fi}}
\def\@array[#1]#2{\setbox\@arstrutbox=\hbox{\vrule
     height\arraystretch \ht\strutbox
     depth\arraystretch \dp\strutbox
     width\z@}\@mkpream{#2}\edef\@preamble{\halign
\noexpand\@halignto
\bgroup \tabskip\z@ \@arstrut \@preamble \tabskip\z@ \cr}%
\let\@startpbox\@@startpbox \let\@endpbox\@@endpbox
  \if #1t\vtop \else \if#1b\vbox \else \vcenter \fi\fi
  \bgroup \let\par\relax
  \let\@sharp##\let\protect\relax
  \@arrayskip\@preamble}
%
%
%
%
\def\eqnarray{\stepcounter{equation}%
              \let\@currentlabel=\theequation
              \global\@eqnswtrue
              \global\@eqcnt\z@
              \tabskip\@centering
              \let\\=\@eqncr
              $$%
 \halign to \displaywidth\bgroup
    \eqnumphantom\@eqnsel\hskip\@centering
    $\displaystyle \tabskip\z@ {##}$%
    \global\@eqcnt\@ne \hskip 2\arraycolsep
         $\displaystyle\arraymode{##}$\hfil
    \global\@eqcnt\tw@ \hskip 2\arraycolsep
         $\displaystyle\tabskip\z@{##}$\hfil
         \tabskip\@centering
    &{##}\tabskip\z@\cr}


\setcounter{footnote}0

\setcounter{footnote}0

\begin{flushright}
ITEP/TH-77/98\\
hepth/9812519
\end{flushright}
\vspace{0.5cm}

\begin{center}

{\LARGE\bf Low energy theorems and the Dirac operator spectral density
in QCD}

\bigskip {{\Large A.Gorsky}   \\
 ITEP, Moscow, 117259, B.Cheryomushkinskaya 25}

\end{center}
\bigskip

\begin{abstract}
We discuss the behaviour of the spectral density of the
massless Dirac operator
at the small eigenvalues and quark masses compatible with the
restrictions
imposed by the low energy theorems in QCD. Sum rule for its derivative
over the quark mass is found.

\end
{abstract}
\bigskip

1. The search for the universal characteristics of  QCD at the strong
coupling
regime remains the important problem to clarify the
structure of QCD vacuum. One of the most important universal
objects is the spectral density of the
massless Dirac operator whose behaviour near
zero eigenvalue provides  the pattern of the
spontaneous symmetry breaking. The Banks-Casher relation \cite{bc}
states that the
density at the origin is the fermionic condensate $ <\bar q q>=-\pi
\rho(0)$,
where
\be
\hat D q=\lambda q \\
\rho(\lambda)=<V^{-1}\sum\delta(\lambda-\lambda_n)>_{A}
\ee
V is the Euclidean volume and averaging over the gluon ensemble is
assumed. Generalization of the Banks-Casher relation for the moments
of the density at the finite volume  has been found in \cite{ls}.
The obvious question concerns the behaviour of the spectral density at
small masses and eigenvalues.
In the perturbation theory density behaves as $\rho(\lambda)=c\lambda^3$

therefore the linear and quadratic terms have the nonperturbative
nature. It would describe the critical behaviour
of the system at zero temperature and possibly fix a universality class.

Let us remark that the universality
properties of the spectral density
allow to apply the matrix model technique to
analyse its behaviour at the finite volume (see \cite{ver} for the
review).

It is natural to relate the characteristics of the spectral density to
other universal objects in QCD. The best candidates are the low energy
theorems for the zero-momentum QCD correlators in the different channels

which amount from the chiral Ward identities \cite{gl,nsvz}. The first
attempt
to get the information about the spectral density behaviour
involved isovector scalar correlator $\int dx<S^i(x) S^j(0)>$ \cite{ss}.

It was claimed that
it yields the term linear in $\lambda$ that   vanishes
for $N_{f}=2$. Recently two other sum rules arising from the
correlators
$$
\int dx<A^i(x)A^j(0)-V^i(x)V^j(0)>
$$
\cite {s} and
$$
\int dx<S^i(x) S^j(0) - \delta^{ij}P^0(x)P^0(0)>
$$
\cite{sm}
were obtained but no additional information
on the spectral density has been extracted.

In this note we discuss the complete set of restrictions imposed by low
energy
theorems for the two point and three point correlators
on the spectral density.
Correlators in the scalar and pseudoscalar channels yield the
information
on the  spectral density itself while those in the vector and axial
channels
are relevant only for the  correlations of the eigenvalues.

2.
Let us consider the correlators in the isovector and isoscalar
scalar and
pseudoscalar channels in the $N_{f}=2$ case. In what
follows we shall discuss only
correlators which are free from the nonuniversal high energy
contributions. The complete list of such low- energy theorems for two ,
three and four point correlators can be found in \cite{gl}.
Corresponding
correlators can be expressed in terms of the spectral density as
follows;

\be
\int dx<\delta^{ij}S^0(x) S^0(0) -P^i(x) P^j(0)>=-\frac{G_{\pi}^2
\delta_{ij}}{M_{\pi}^2}+ \delta_{ij}\frac{B^2}{8\pi^2}(l_3-4l_4+3)= \\
-\delta_{ij}(\int \frac{8m^2\rho(\lambda,m)}{(\lambda^2+m^2)^2}-
4m\int \frac{\partial_{m}\rho(\lambda,m)}{(\lambda^2 +m^2)})
\ee

\be
\int dx <S^i(x) S^j(0) - \delta_{ij} P^0(x) P^0(0)>= -\delta_{ij}
8B^{2}l_{7}
=- \\ \delta_{ij}(\int
\frac{8m^2\rho(\lambda,m)}{(\lambda^2+m^2)^2}-4\frac{<\int
dx
Q(x)Q(0)>}{m^2V})
\ee

\be
\int dx <P^3(x) P^0(0)>=\frac{G_{\pi}\tilde G_{\pi}}{M_{\pi}^2} = \\
8(m_{u}-m_{d})m\int \frac{\rho(\lambda,m)}{(\lambda^2 +
m^2)^2}-\frac{4(m_{u}-m_{d})<\int dx Q(x)Q(0)>}{m^3V}
\ee
where $G_{\pi}= 2F_{\pi}B=\frac{F_{\pi}m_{\pi}^2}{m};
\tilde G_{\pi}=-(m_{u}-m_{d})\frac{4B^2 l_{7}}{F_{\pi}}$
and $ Q(x)$ is the topological charge density.
Sum rule arising from the last low energy theorem exactly coincides with

the one coming from correlator (3). Low energy constants $l_3,l_4$
behave as $log m$ at small quark masses, while the constant $l_7$
contains no chiral logarithms \cite{gl}.

We can add here low energy theorem for three point correlator
\be
\int dxdy<S^{0}(x)P^{i}(y) P^{k}(0)>=\frac{\delta_{ik}
G_{\pi}^3}{M_{\pi}^4
F_{\pi}}-
\delta_{ik}\frac{B^2G_{\pi}F_{\pi}}{8\pi^2M_{\pi}^2}(l_3-4l_4+3)=\\
\delta _{ij}\frac{d}{dm}\int \frac{2\rho(\lambda,m)}{\lambda^2 +m^2} .
\ee
It appears that the sum rules resulted  from the low energy theorems (2)

and (5) are identical.

The last low energy theorem  which is potentially
important amounts from the four
point pseudoscalar correlator
\be
\int dxdydz<P^i(x)P^j(y)P^k(z)P^l(0)>=-\frac{G_\pi^4}
{F_\pi^2 M_{\pi}^8}(\delta_{ij} \delta_{kl} +
\delta_{ik}\delta_{jl}+\delta_{il}\delta_{jk})
(-\frac{M_{\pi}^2}{F^2} +\frac{M_{\pi}^4(24l_4-9)}{96\pi^2F^4}+...) = \\

(\delta_{ij} \delta_{kl} +
\delta_{ik}\delta_{jl}+\delta_{il}\delta_{jk})\int
\frac{4\rho(\lambda,m)}
{(\lambda^2+m^2)^2}+R_{ijkl}.
\ee
Unfortunately it is more difficult to extract the information
from the four point correlator since there is the contribution
$R_{ijkl}$ corresponding to the diagram with
at least two fermionic loops
which can,t be expressed in terms of the spectral density .
Moreover the sum rules are sensible to the two loop contribution
to the four point correlator in the chiral theory which is
unknown at a moment. Therefore there are only two
rigorous independent
sum rules for the spectral density.

Combination of the sum rules (2) and (3)
yields the following model independent
sum rule for the mass derivative
of the spectral density
\be
m\int \frac{4\partial_{m}\rho(\lambda,m)}{(\lambda^2 +m^2)} =\\
\frac{B^2}{8{\pi}^2}(l_3-4l_4+3)- \frac{G_{\pi}^2}{M_{\pi}^2}+
8B^2 l_7 +\frac{4<\int dx Q(x)Q(0)>}{m^2 V}.
\ee

3.
Turn now to the discussion of the restrictions imposed
by the sum rules on the
behaviour of the density near the origin. We
would like to look for the following
anzatz
\be
\rho(\lambda,m)=\rho(0)+ c_1{\lambda} log{\lambda} +c_2\lambda+ c_3m+
c_4mlogm+c_5mlog^2 m+O(m^2,\lambda^2,m\lambda),
\ee
where $c_i$ are the constants to be found.
Let us first make an assumption that the spectral density has to be
regular
if $\lambda\rightarrow 0$ or $m \rightarrow 0$ so there
are no terms like $(\frac{\lambda}{m})^n$ or $mlog\lambda$ in the
expansion above.
It appears that this anzatz is
consistent with the low-energy theorems for the correlators
which are free from the high energy contributions.

Apart from the sum rules above we assume two additional model
independent restrictions on the spectral density. First we  use
the universality of the $mlogm$ correction to the chiral condensate
\cite{nsvz}. Secondly there is   unambiguous statement that
there are no $logm$ contributions to the correlator $<S^jS^i>$. This
fact has been already used  to show that $c_2=0$ \cite{ss}.

Consider first correlators
(2),(5) which give rise the identical sum rules. The leading $m^{-1}$
singularity immediately comes from $\rho(0)$ term but the matching
of the $logm$ term appears to be a subtle point. It is easily seen that
the constants $c_4,c_1$ don,t contribute hence one has to assume
that  $c_5\neq 0$.

The correction to the condensate looks as
\be
<\bar q q>_m=<\bar q q >_0(1-\frac{3M_{\pi}^2logM_{\pi}^2}{32 \pi^2
F_{\pi}^2}+
...)
\ee
and it is supposed that there is no $mlog^2m$  term.
Hence using the Banks-Casher formula for the condensate we can claim
that to cancel $mlog^2m$ correction we have to assume that $c_1 \neq 0$.

However this term in the spectral density yields the divergence in the
integral at the UV region rising the question on the
substraction of the perturbative contribution.
The proper version of this proceedure
which would allow to make a prediction
for $c_1,c_4$ deserves further investigation.

To discuss the restriction on $c_3$ we have to expand the topological
susceptibility up to the second order in the quark mass
\be
\frac{<\int dx
Q(x)Q(0)>}{V}=\frac{mBF_{\pi}^2}{2}+dB^2m^2+... ,
\ee
hence the  coefficient d enters
our sum rules. If we substitute the expression for the spectral density
and
assume that all logarithmic contribution cancel the model independent
relation  arises
\be
c_3=-8l_7 +2d.
\ee
Note
that the analogous relation has been considered in \cite{sm} but the
term in
the lhs has been missed.

Since the constants $c_1$ and $c_5$ can,t vanish simultaneously
the $m^2logm$ and $m^2log^2m$ terms   have to be manifest in the
expansion of the topological susceptibility. While the $logm$
term can be attributed to the first correction to the
condensate the origin of the $log^2m$ term is unclear.
Certainly it is desirable to consider the $N_{f}>2$
case when the dependence of the topological susceptibility
on the quark masses is more transparent.

Recently the behaviour of the spectral density at small quark
masses with arbitrary number of flavours has been found within
the matrix theory approach \cite{otv}. It was claimed that the
singular terms are present in the diffusive domain where the
spectral density manifests the $mlog\lambda$ contribution. It can
be shown that such term is consistent with the sum rule found
above. However it is not clear how possible singularity agrees
with the nonzero quark condensate.

4.
In this note we considered the behaviour of the
Dirac operator spectral  density
around the origin. It appeared that the low energy theorems impose
strong restrictions on it but don,t determine it unambigously. It
would be  crusual to derive the complete list of the two loop
contributions to the
off shell correlators in the chiral theory
since this information provides the additional restrictions on
the linear terms in the spectral density as well as on the
$\lambda^2$ terms. We have shown that the nonsingular spectral density
manifests unexpected $mlog^2m$ term which possibly can be
attributed to the renormalization of the mass in the strong
external field similar to the QED renormalization of the mass
\cite{gus}.
Otherwise some singular terms \cite{otv} in the spectral density
have to be admitted.

This work is partially motivated by an attempt to develop the
interpretation of the spectral density of the Dirac operator
within the brane approach. This issue was addressed for the N=1
SUSY theory in \cite{gor} and
quark masses fix the positions of D6 branes representing
the fundamental matter in the "momentum" space.
Therefore to clarify the brane picture it is necessary
to elaborate the $N_f$ dependence of the spectral density as well
the the case of the generic mass matrix. These questions will
be discussed elsewhere.

I am grateful to H.Leutwyler for the hospitality in ITP at
Bern University  and discussions
which initiated this work. I would like to thank A.Smilga for
the useful comments and D.Toublan
for bringing reference \cite{otv}
to my attention. The work was supported in part by grants
INTAS-96-0482   and RFBR 97-02-16131.

\end{document}